# Solar coronal loop dynamics near the null point above active region NOAA 2666

Boris Filippov

Pushkov Institute of Terrestrial Magnetism, Ionosphere and Radio Wave Propagation of the Russian Academy of Sciences (IZMIRAN), Troitsk, Moscow 108840, Russia
(e-mail: bfilip@izmiran.ru)

**Abstract**
We analyse observations of a saddle-like structure in the corona above the western limb of the Sun on 2017 July 18. The structure was clearly outlined by coronal loops with typical coronal temperature no more than 1 MK. The dynamics of loops showed convergence toward the centre of the saddle in the vertical direction and divergence in the horizontal direction. The event is a clear example of smooth coronal magnetic field reconnection. No heating manifestations in the reconnection region or magnetically connected areas were observed. Potential magnetic field calculations, which use as the boundary condition the *SDO*/HMI magnetogram taken on July 14, showed the presence of a null point at the height of 122" above the photosphere just at the centre of the saddle structure. The shape of field lines fits the fan-spine magnetic configuration above NOAA 2666.

## 1 Introduction

Magnetic null points in the solar atmosphere play a great role in the study of fast and energetic phenomena observed on the Sun. They are assumed as favorable places of magnetic energy release and conversion into other forms due to reconnection of field lines (Giovanelli, 1946; Dungey, 1958; Severnyi, 1958; Parker, 1957, 1973; Sweet, 1958; Petschek, 1964; Syrovatskii, 1966, 1971; Sonnerup, 1970; Somov, 1992; Priest & Titov 1996; Priest and Forbes, 2000). The individual field lines near a saddle-type (X-type) null point have the shape of hyperbolas, except the field lines located in some plane called a fan surface, which separates two families of lines, and two oppositely directed straight lines called a spine (Priest and Pontin 2009). One of the simplest and probably most common situations is the appearance of a null point above an inclusion of parasitic polarity inside a vast unipolar region (Lau & Finn 1990; Priest et al. 1994; Antiochos 1998; Pariat et al. 2009; Pontin et al. 2013). The characteristic shape of field lines near the null point delineated by coronal loops provides the means to detect null points in the corona, since the magnetic field in the solar corona is so far almost inaccessible to measurement.

Saddle-like structures in the corona were observed in soft X-ray and extreme ultraviolet (EUV) ranges. Some of them are observed in projection on the disc (Tsuneta 1996; Filippov, 1999a, Sun et al. 2014; Xue et al. 2016, Li et al. 2016). They can be, however, saddle-like only in two dimensions with a non-zero third component. Other saddles, especially observed above the limb, are definitely three-dimensional (3D) nulls (Filippov, 1999b; Su et al. 2013; Freed et al. 2015).

It is possible also to search null points in the corona using different methods of a photospheric magnetic field extrapolation (Schrijver and Title 2002; Longcope, Brown, and Priest 2003, Régnier, Parnell, and Haynes 2008; Longcope and Parnell 2009; Cook, Mackay, and Nandy, 2009; Baumann, Galsgaard, and Nordlund 2013; Platten *et al.*, 2014; Freed et al. 2015 ; Edwards and  Parnell 2015). Since these calculations reveal a lot of null points above a comparatively complicated photospheric field, the distribution of null points with height, latitude, and cycle were studied (Cook, Mackay, Nandy 2009; Edwards and  Parnell 2015) as well as their association with eruptive events (Ugarte-Urra et al. 2007; Barnes 2007), flares (Wang and Wang

1996; Fletcher et al. 2001; ) and coronal jets (Shibata K., et al., 1992; Filippov, Golub & Koutchmy 2009; Moore et al. 2010; Sterling et al. 2015).

High resolution EUV images with a high cadence provided by the Atmospheric Imaging Assembly (AIA; Lemen et al. 2012) on board the *Solar Dynamics Observatory* (*SDO*; Pesnell et al. 2012) tempted to find direct evidence of coronal reconnection in the changing of the loop connectivity. Su et al. (2013) presented observations that provided solid visual evidence of magnetic reconnection producing a solar flare. Cool loops inflow into the reconnection region, merge and disappear, while hot reconnected loops outflow in the perpendicular direction.

Sun et al. (2015) reported evidence of reconnection below an ascending coronal dark cavity, which is assumed to be the cross-section of a helical magnetic flux rope. With the rise of the cavity, the underlying loops, with anti-parallel directions of the magnetic field within, gradually approach each other. They form an X-shaped structure, and immediately after the disappearance of the cool loops a hot region appear near the structure. Flare loops then start to rise below the reconnection region.

Li et al. (2016) and Xue et al. (2016) studied observations of magnetic reconnection stimulated by erupting solar filaments. X-shaped structures in these cases were observed on the disc. Li et al. (2016) observed the encounter of an erupting filament with nearby coronal loops in *SDO*/AIA EUV channels. An X-type magnetic configuration was formed as a result of interaction. Bright current sheets appear at the interfaces of the filament with the loops. Reconnection changed connection of the filament. Xue et al. (2016) used observations with the Chinese New Vacuum Solar Telescope in Hα line supplemented by *SDO*/AIA EUV observations. They studied reconnection between chromospheric fibrils and threads of an erupting filament. Inflows into X-shaped structure and outflows from it were observed. Newly formed loops demonstrate the change of connectivity.

In all cited studies reconnection was fast. It was caused or accompanied by eruptive and flaring phenomena. Thermal effects, heating of reconnected loops and footpoint brightenings, were considered as additional arguments in favour of reconnection evidence.

In this paper, we analyse observations of a saddle structure in the corona above the western limb of the Sun. The structure was clearly outlined by coronal loops with typical coronal temperature no more than 1 MK. We study the dynamics of loops showing convergence toward the centre of the saddle in the vertical direction and divergence in the horizontal direction. The event is a clear example of smooth coronal magnetic field reconnection. No heating manifestations in the reconnection region or magnetically connected areas were observed.

**2 Data sets**

The main source of information about the studied event was observations in EUV provided by the AIA instrument on board *SDO*. AIA takes full–disc images of the Sun in several EUV, ultraviolet, and continuum wavebands with the spatial resolution of $0.6''$ per pixel. We used predominantly images obtained in the 171 Å channel, because in this channel coronal loops are visible with the most contrast. However, images from other channels were also used to see temperature variations in the region where reconnection of coronal loops seems to occur. Data taken by the Helioseismic and Magnetic Imager (HMI; Schou et al. 2012) on board *SDO* were used to analyse the photospheric magnetic field distribution and as a boundary condition for the solution of the Newman boundary-value problem. This solution provides the shape of magnetic field lines in the corona in the potential-field approximation.

## 3 Observations of the coronal loop evolution

On 2017 July 18, coronal loops observed at the western limb of the Sun showed remarkable dynamics near the coronal structure resembling the appearance of a saddle. The loops, presumably according to their temperature, were better visible in the *SDO*/AIA 171 Å channel. The saddle structure was located above newly emerged active region NOAA 2666 and can be recognized on the disc for several previous days (Fig. 1). More developed and older active region NOAA 2665 was situated nearly symmetrically relative the equator in the southern hemisphere (Figs 1 c and f). Transequatorial loops connected a big positive sunspot in NOAA 2665 with negative polarities in NOAA 2666 (Fig. 1d). The height of the saddle above the limb on July 18 was 50". Since active region NOAA 2666 did not reach the limb by this time, the height of the saddle above the positive polarity in AR 2666 was about 115".

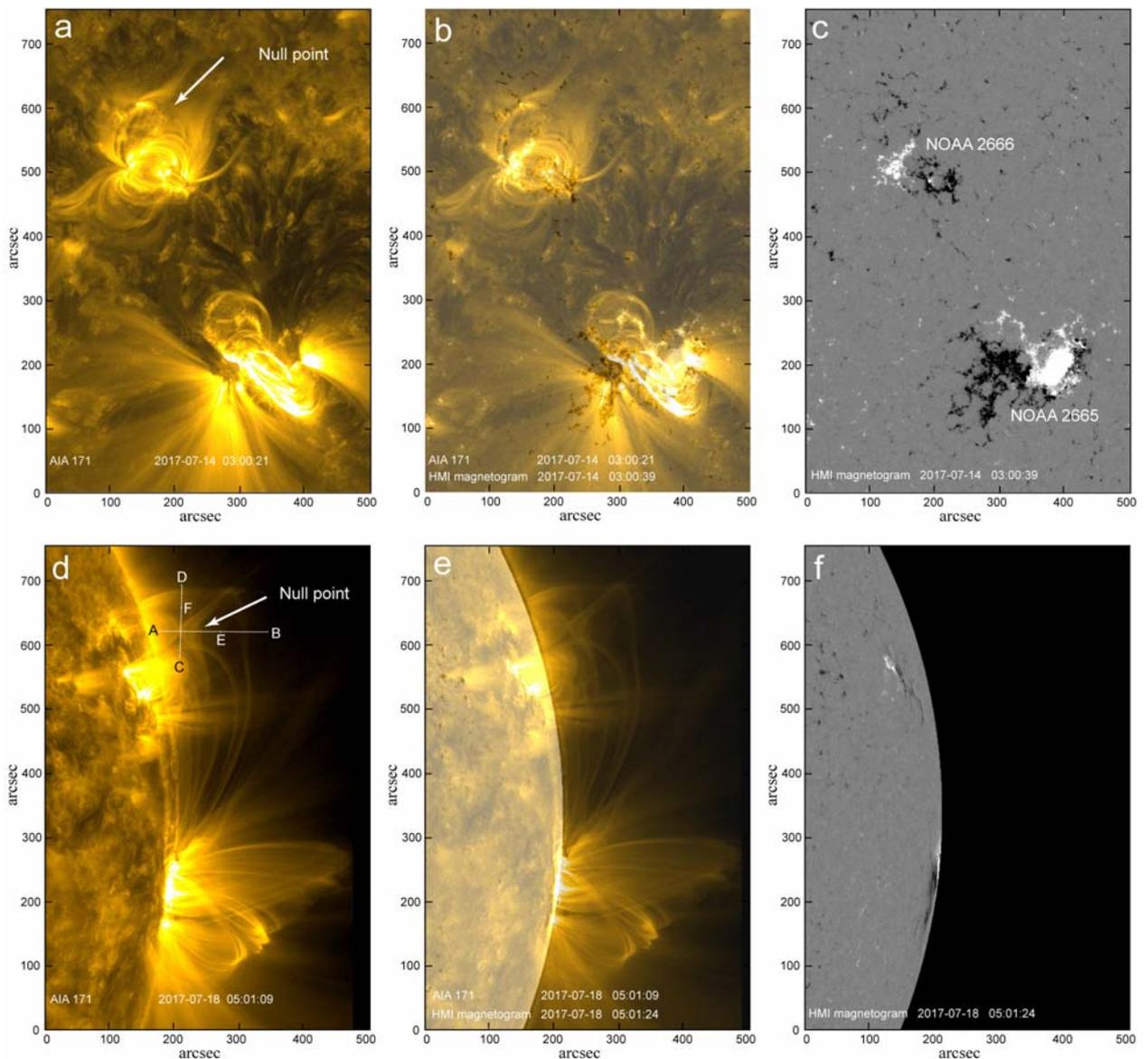

**Figure 1.** Coronal loops observed in the 171 Å channel of the *SDO*/AIA instrument on 2017 July 14 and 18 (left panels) and *SDO*/HMI magnetograms of the same regions (right panels). In the central panels, the coronal images are superposed on the magnetograms. The white arrow points to the saddle structure indicating the coronal null point. (Courtesy of the *SDO*/AIA and *SDO*/HMI science teams.)

The saddle-like shape of the coronal structure hints on the presence of a magnetic-field null point, which is surrounded by hyperbolic field lines. Null points are preferable places for field-line reconnection. During reconnection, field lines approach the null-point from two opposite directions. They change their connectivity continually and continuously as they pass through the non-ideal region around the null-point where magnetic diffusion is possible (Priest et al 2003; ponin et al 2013). Reconnected field lines go away from the null-point perpendicularly to the initial direction. Similar dynamics show coronal loops in the vicinity of the saddle structure on July 18. Fig. 2 presents a number of snapshots taken with the *SDO*/AIA 171 Å filter (see also movie1). Images are processed using the unsharp-mask filter and contrasted. One system of rather small loops rises from lower heights to the centre of the saddle. Another loop system consisting of long transequatorial loops descends from above. The curvature of these loops within the saddle and nearby is directed upward, which is atypical for coronal loops. There is a dark gap between the two loop systems. About 04:30 UT both systems touch each other and some features diverging horizontally from the centre of the saddle can be recognized.

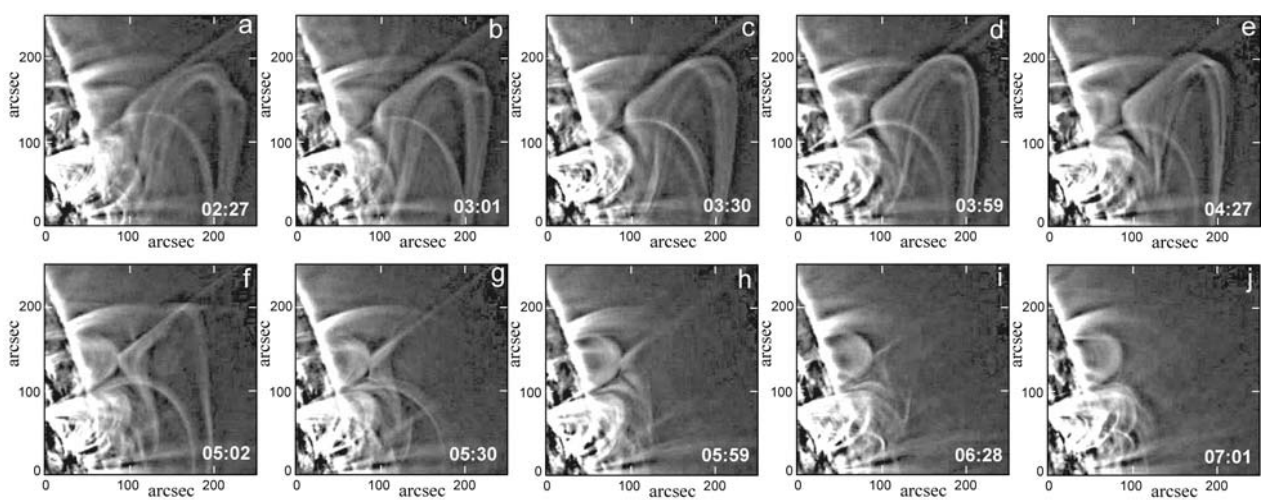

**Figure 2.** Evolution of coronal loops observed in the *SDO*/AIA 171 Å channel near the null point on 2017 July 18. (Courtesy of the *SDO*/AIA science team.)

The dynamics of the loops is visible more clearly in time-slice plots made for two perpendicular slit positions shown as the lines A-B and C-D in Fig. 1(d). For the vertical slit direction (Fig. 3a), bright strips incline to the null point. Blue dashed lines show some approaching to the null point loops. The velocities are in the range of 4 – 5 km/s. The diverging motion becomes visible in Fig. 3(b) after 04 UT. Two loops most recognizable in Fig. 3(b) diverge rather slowly with velocities of 2.3 and 3.4 km/s. However, there are traces of short track after 05 UT that have greater inclinations. They are recognized better in Fig. 4(b) showing loop evolution near the null point in more detail. There are a number of short tracks with different inclinations. Two most prominent loops move with velocities of 4.5 and 5.5 km/s. The inclination of faint tracks in the right-side part of Fig. 4(b) is greater, so they can be considered as moving faster.

The detailed evolution of loops near the null point is visible in movie2. Upward concaved long loops approach to the null point from above in the radial direction, while new vertical loops appear sequentially and move away from the null in the horizontal direction. Several horizontal loops are seen to move upward from the null point in the right-hand part of the movie2 frame. Comparison with movie1 showing a larger field of view convinces that these loops are far beyond the null point position to the west of it. The motion of these loops is possibly related to some eruptive activity originated to the north-west from the studied region. They show an interesting large-scale dynamics, but they are rather faint and do not related to the loop evolution near the null point we discuss. Fig. 5 shows three difference images demonstrating the transformation of one particular loop near the null point. The V-shaped dark structure above the null approaches to it from above and disappears, while a new also V-shaped white structure appears and moves away from the null in the perpendicular direction.

We do not observe any thermal-excess manifestations in the region. Fig. 6 presents snapshots at 05:01 UT in the other EUV channels of the *SDO*/AIA instrument. Most of them (except the 304 Å channel) show emission of plasma hotter than plasma emitting within 171 Å band. The saddle structure can be vaguely recognized only in the 131 Å channel, which is also sensitive to rather cool plasma (Fig. 6b). The other images do not show even faint traces of the structure.

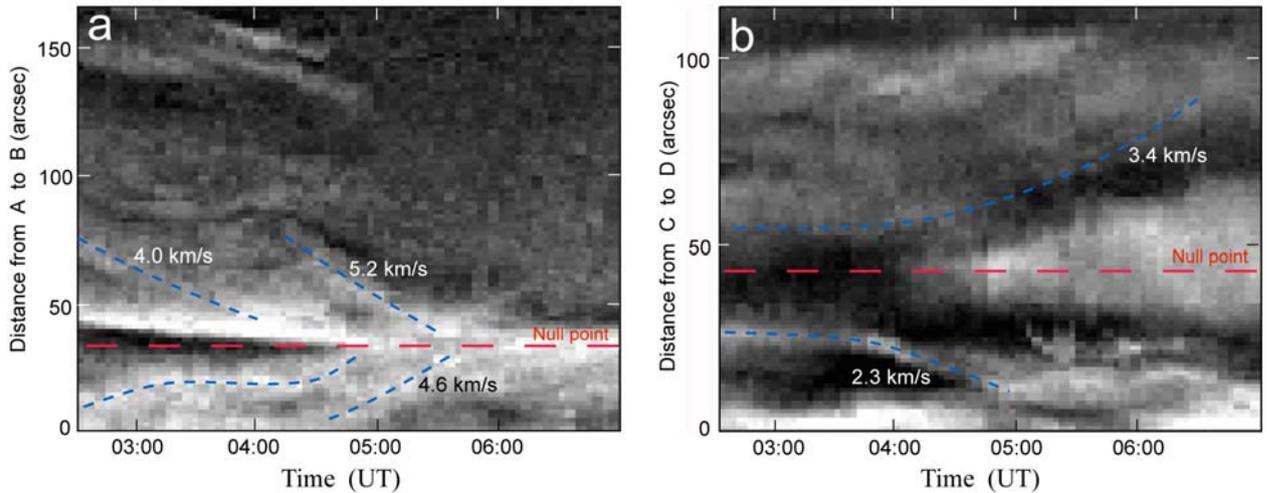

**Figure 3.** Time-slice plots for the slit positions shown as the lines A-B and C-D in Fig. 1(d). Red dashed lines show the centre of the saddle, the presumable location of the null point. Blue dashed lines indicate trajectories of most prominent loops.

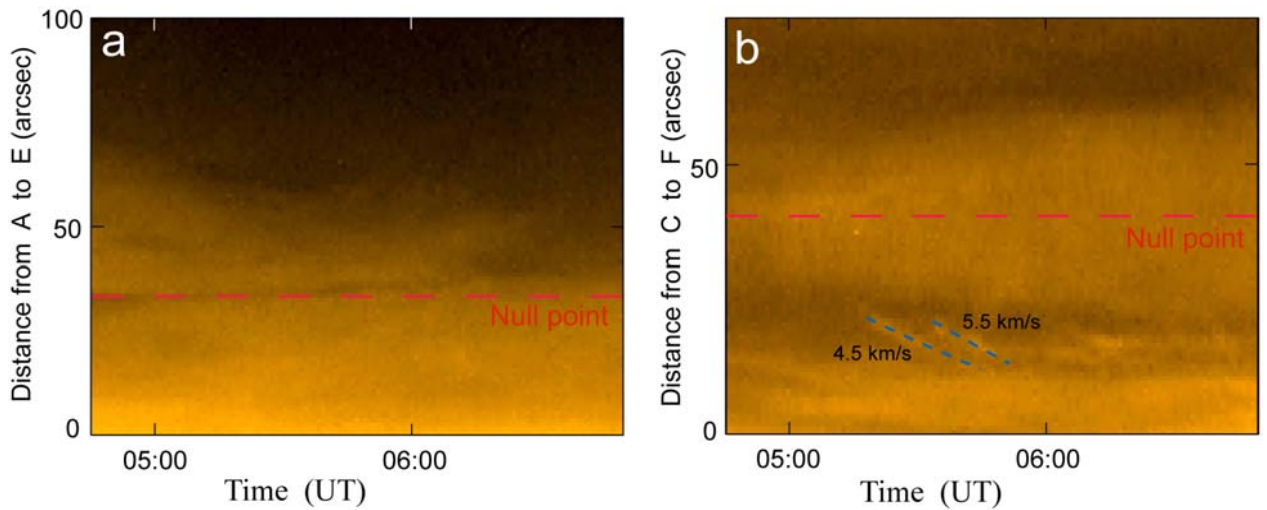

**Figure 4.** Time-slice plots for the slit positions shown as the lines A-E and C-F in Fig. 1(d). Red dashed lines show the centre of the saddle, the presumable location of the null point. . Blue dashed lines indicate trajectories of most prominent loops.

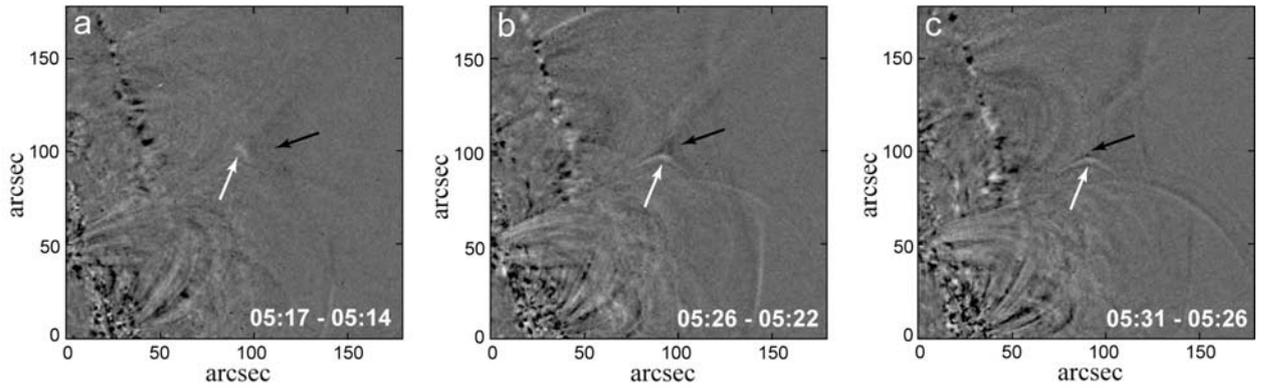

**Figure 5.** Difference images of coronal loops observed in the *SDO*/AIA 171 Å channel near the null point. Black arrows point to dark structures visible in preceding images. White arrows indicate structures that appear in subsequent images. (Courtesy of the *SDO*/AIA science team.)

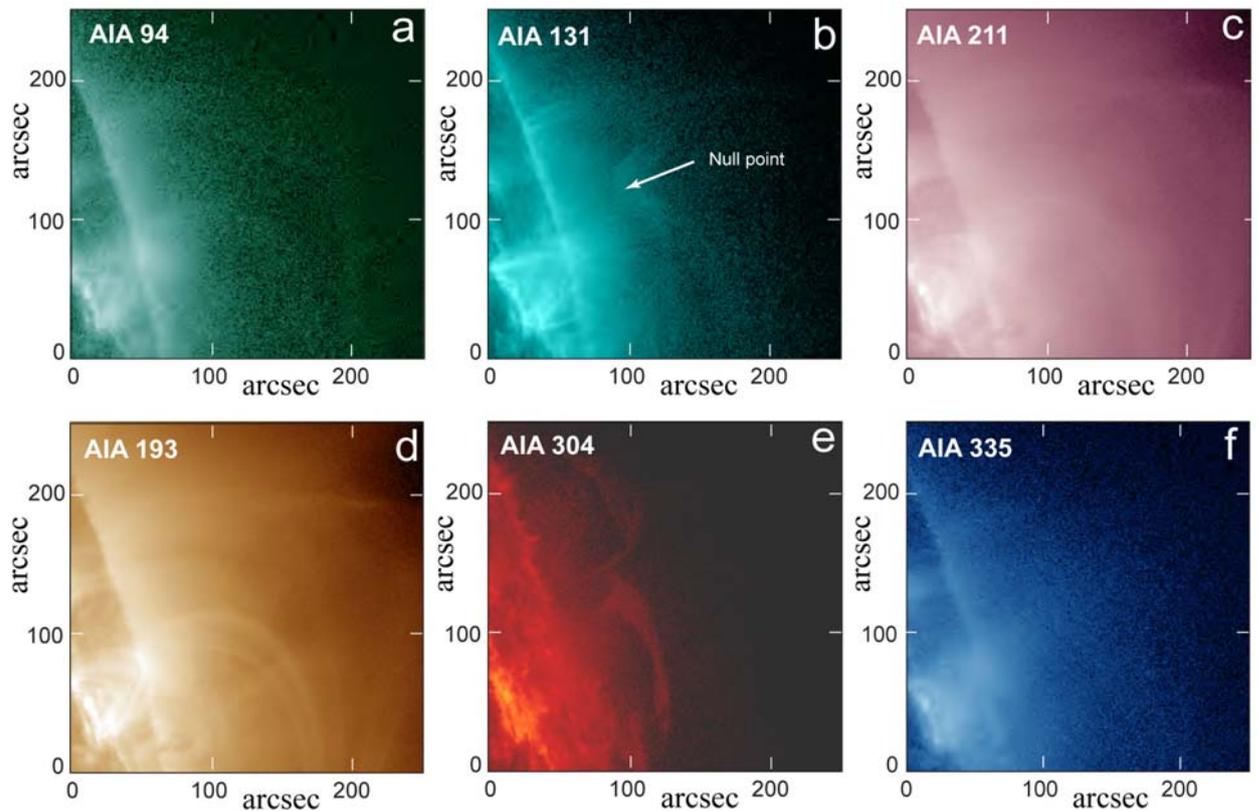

**Figure 6.** *SDO*/AIA images in different channels on 2017 July 18 at 05:01 UT. (Courtesy of the *SDO*/AIA science team.)

**4 Coronal magnetic field structure**

The saddle structure formed by coronal loops is a strong argument in favour of the presence of real 3D magnetic-field null point in the corona. However, the existence of the null point should be confirmed by photospheric magnetic field extrapolation. For instance, the observed structure may be saddle-like only in two dimensions presented in an image. An example of such structure is visible in Fig. 1(a) between two active regions. It is located on the line connecting major negative polarities of the two active regions. The horizontal components vanish at the centre of the saddle structure, but the third, vertical, component is non-zero. This may be guessed on the basis of the photospheric field distribution and confirmed by magnetic field extrapolation.

Active region NOAA 2666 emerged on July 12 within a large area of predominantly negative polarity (Fig. 7). Thus, the emerged positive polarity is surrounded by dominating negative polarity. This photospheric field distribution inevitably leads to the appearance of a null point in the corona above the minor polarity (Lau & Finn 1990; Priest et al. 1994; Antiochos 1998; Pariat et al. 2009). However, it should be confirmed by coronal field calculations based on photospheric field measurements. The simplest method is to use the current-free approximation for coronal field (Schmidt 1964; Altschuler & Newkirk 1969; Levine 1975; Adams & Pneuman 1976; Filippov and Den 2001).

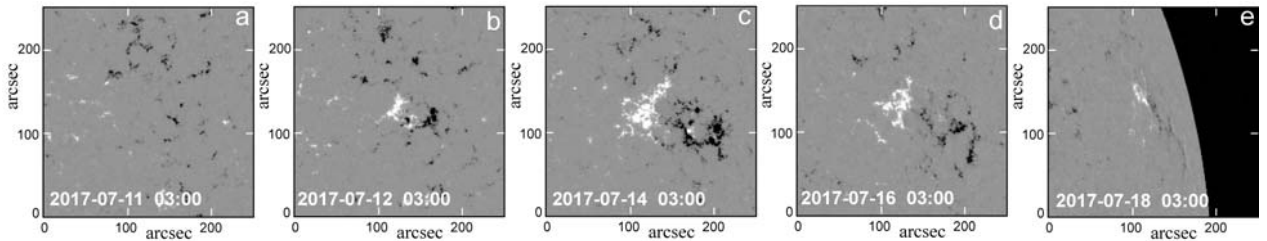

**Figure 7.** *SDO*/HMI magnetograms from July 11 to 18 showing the emergence of the positive magnetic flux in NOAA 2666. (Courtesy of the *SDO*/HMI science team.)

On July 18 the region was too close to the western limb. Magnetic field measurements were uncertain and noisy in this area due to an acute angle between the photospheric surface and the line-of-sight. We should choose a magnetogram taken on one of the previous days when the region was on the disc not far from the central meridian and assume that the large-scale structure of the field does not change significantly during this time interval. We used the *SDO*/HMI magnetogram taken on July 14 at 03:00 UT, which is shown in Fig. 1(c). Fig. 8 presents results of coronal field calculations in current-free approximation with the use of the Green's function technique (see *e.g.* Filippov 2013 and references therein). The general topology of the coronal field is shown in Fig. 8(a). Two field lines converging above the positive polarity indicate a fan-spine magnetic configuration. A null point may be expected below the most curved sections of these lines.

We looked for null points using the necessary condition for the existence of a null point within a grid-cell: all three field components should change the sign at each corner of the cell (Haynes and Parnell, 2007). We used the grid with the size of cells of 8" and found eight adjacent suspicious cells at a height of about 120". Since we do not need the precise position of the null point, we demonstrate its existence in Fig. 8(b) by showing the coincidence of zero-points of the horizontal and vertical field components at the height of 122". Zero-lines of the vertical field (polarity inversion lines) are shown as red contours. The small red circle with the coordinates (270", 480") is located at the centre of the area with diverging in all directions arrows. According to continuity of the field, the horizontal component vanishes in this place as well as the vertical component. So, this is a real 3D null point. The height of the calculated null point is very close to the measured height of the saddle centre in the *SDO*/AIA images.

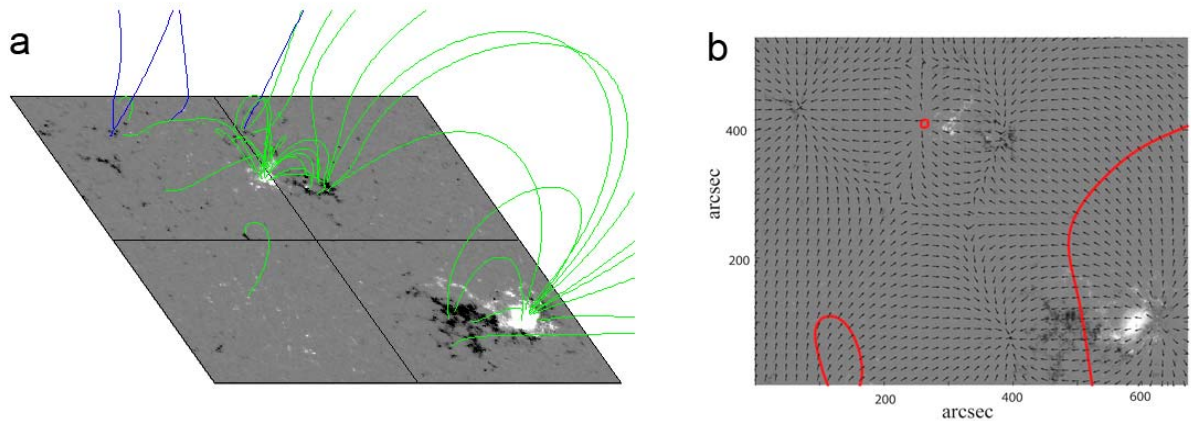

**Figure 8.** Potential magnetic field lines above photospheric magnetogram of the active regions NOAA 2665 and 2666 (left) and directions of the horizontal magnetic field (arrows) and polarity inversion lines (red lines) for the height of 122" superposed on the magnetogram (right). Green lines represent field lines starting and ending in the photosphere, while not all their length may be shown. Blue lines show the field lines ending at the upper boundary of the calculation domain.

## 5. Discussion and conclusions

We found a saddle-like structure formed in the corona by coronal loops visible in 171 Å *SDO*/AIA images on 2017 July 18. The structure changes slowly during 4 h. The separatrices of the saddle are nearly perpendicular to each other as it is expected in current free field and are inclined to the solar surface and radial direction by the angles about 20°. In 2D images, we see a projection of a real 3D structure on the plane of the sky. Time sequence of images and the photospheric magnetic field distribution suggesting some field-line connectivity help to disentangle the observed coronal loop picture, however some ambiguity remains in our conclusions. We suspect that the observed above the limb saddle-like structure relates to 3D coronal magnetic-field null point in contrast to more ubiquitous 2D saddle-like structures visible on the disc (as e.g. can be recognized in the central part of Fig. 1(a) between two active regions). Potential magnetic field calculations, which use as the boundary condition the *SDO*/HMI magnetogram taken on July 14, showed the presence of a null point at the height of 122" above the photosphere just at the centre of the saddle structure. The shape of field lines fits the fan-spine magnetic configuration above NOAA 2666.

Magnetic flux tubes become visible in coronal images due to inhomogenities of plasma density and temperature. Since these inhomogenities more easily spread along the direction of magnetic field, sets of thin individual threads are observed in the corona as coronal loops. They usually called loops because a loop is the most typical shape of their axes, but they also can have a shape of segments of straight line owing to a large size of a loop or opening into the interplanetary space. The fan-spine magnetic configuration has more o less regular cylindrical symmetry. The saddle structure is conspicuous because some flux tubes are more noticeable. One reason for better visibility is the location of the great part of the tube in the sky plane; another reason is the distribution of the main photospheric magnetic sources in this region along the meridian (also close to the sky plane). However, at different times different loops become more prominent.

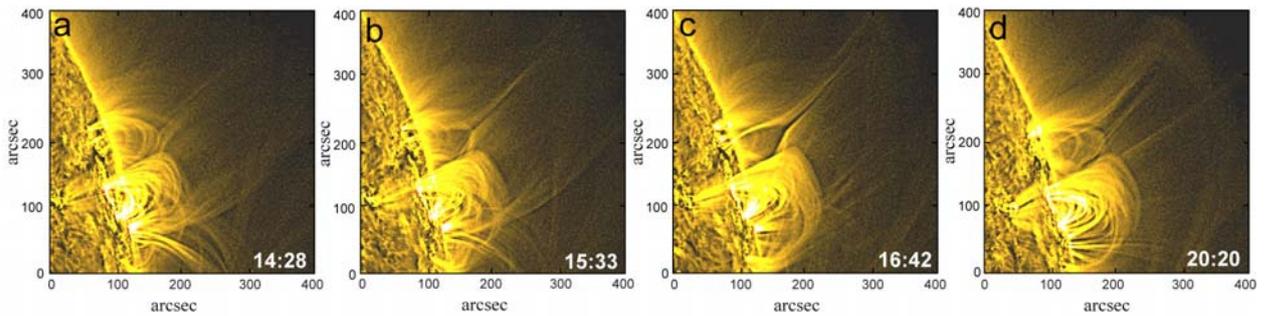

**Figure 9.** *SDO*/AIA images in the 171 Å channel on 2017 July 18 showing the outer part of the fan-spine configuration. (Courtesy of the *SDO*/AIA science team.)

Before the reconnection event only two systems of hyperbolic lines with loops horizontal near the saddle centre are clearly visible (Fig. 2). The loops move slowly toward each other and meet at the saddle centre. If coronal loops represent magnetic field tubes filled with coronal plasma, the observed picture visualizes magnetic field line reconnection at a null point. Potential magnetic field calculations confirm the presence of the null point at the place where the saddle structure is observed. Field lines approach the reconnection region from below and from above. After reconnection they should form field lines directed vertically near the null point. Such lines moving away from the saddle centre really appear in *SDO*/AIA images (Fig. 5, movie2) and can be recognized in the time-slice diagram (Fig. 4b). While the structure of field lines near a 3D null point is different from the field-line pattern in a 2D case and their dynamics during reconnection also quite different (Priest and Titov 1996; Priest et al 2003; Priest and Pontin 2009; Pontin 2011), the observed loop dynamics is similar to the classical 2D reconnection behavior.

The saddle structure disappears after 07 UT (Fig. 2j) and can be recognized again after 14 UT (Fig. 9a). However, in this case the lower loops simply fade away, while a pair of bright loops outlines the outer part of the fan-spine configuration with the locally open flux (Fig. 9b,c). Below the separatrix dome surface, there are no bright loops during this time, and all region of closed flux is dark. The outer spine is also presented by narrow dark gap between the two straight sections of the locally open loops. In fact, they are connected to the large positive source in the southern active region. Later, closed loops appear again below the dome separatrix (Fig. 9d).

Velocities of the loop motion near the null point are rather slow, of the order about 2 – 6 km/s. Thus, this is an example of slow magnetic reconnection. Possibly, because of the low rate of reconnection, there are no thermal manifestations of the process in *SDO*/AIA channels showing hotter plasma. The observed velocities are very slow comparative to ones reported in other studies of coronal reconnection (Su et al. 2013; Sun et al. 2016; Xue et al. 2016), however Sun et al. (2015; 2016) reported the velocities of the inflows varying from 0.1 to 3.7 km/ s in reconnection in the wake of an erupting flux rope and the inflow velocity in a range of 1–10 km/ s in another eruptive event. In contrast to events studied by Su et al. (2013), Sun et al. (2015; 2016), and Xue et al. (2016), the outflow velocity in our event is not much greater than the inflow velocity. The inflow velocity is of the same order as reported by Sun et al. (2015; 2016), while the outflow velocity is lower by nearly an order of magnitude. It should be noted that in our event plasma outflows into the reconnection region in the vertical direction and outflows in the horizontal direction. In the events studied by Su et al. (2013), Sun et al. (2015; 2016), and Xue et al. (2016) the situation is opposite. All these events were more dynamic and X-typed structures seemed to be temporary. They were formed by interaction of an eruptive flux rope with the pre-existing coronal field or by rapidly evolving coronal loops. The event on 2017 July 18 happened in a presumably stable magnetic configuration, which existed at least for several days. Possibly slow reconnection processes occurred permanently in both directions (with vertical inflows and horizontal inflows) but the most favorable conditions for observations on July 18 reveal one of episodes with the most clearness and detail.

The observed configuration is very similar to the model with a slightly inclined spine analysed by Pontin et al. (2013), if all polarities change signs. Of course, the real configuration on the Sun is more complicated than the model. There is strong influence of the remote big active region NOAA 2665, to which all locally open field lines are connected. Thus, the spine above the null deviates first to the north but at greater heights turns to the south to the positive polarity of NOAA 2665 (Figs 1d, 9). However, the observed loop dynamics is nearly the same as Pontin et al. (2013) showed in their Fig. 5 as an example of the spine-fan reconnection, if the direction of time (and therefore the magnetic field changes) is opposite.

Photospheric *SDO*/HMI magnetograms show that the magnetic-field distribution changes slightly from day to day. Therefore, the coronal field topology should be invariable and the saddle should be observed for a long time. Indeed, the saddle structure can be recognized on several previous days on the disc, however the observation perspective was not as favorable as it becomes when the saddle appear above the limb. It should be noted that even in this favorable position the saddle is not clearly visible all the time. The position of a particular field line near a null point is very sensitive to little changes of the boundary conditions. The visibility of a flux tube depends on density and temperature of plasma filling it. Inhomogeneity of plasma parameters, on one hand, allows one to see the structure of the coronal magnetic field, but, on the other hand, show only some isolated tubes with plasma that emits in a range in which the filter in use is sensitive. Other parts of the coronal field might be obscure.

**Acknowledgements**
The author thanks the *SDO*/AIA and *SDO*/HMI science teams for the high-quality data supplied and A. Tlatov for providing a code for field-line calculations and visualization. The images and movies were obtained using the ESA and NASA funded Helioviewer Project.